\documentclass[10pt, a4paper, twocolumn, conference]{IEEEtran}
\usepackage{amsmath,graphicx,subfigure,epstopdf,multirow,color,flushend,cite,url,fancyhdr}
\IEEEoverridecommandlockouts

\begin{document}
\pagestyle{fancy}
\lhead{This article has been submitted for possible publication at the IEEE ICC 2013}

\title{Opportunistic Relaying in Wireless Body Area Networks: Coexistence Performance}

\author{\IEEEauthorblockN{Jie Dong, David Smith}
\IEEEauthorblockA{National ICT Australia (NICTA)$^\dag$ and The Australian National University\\
(e-mail: Jie.Dong@nicta.com.au)
\thanks{$^\dag$NICTA is funded
by the Australian Government as represented by the Department of
Broadband, Communications and the Digital Economy and the Australian
Research Council through the ICT Centre of Excellence program.}}}

\maketitle

\thispagestyle{fancy}
\lhead{This article has been submitted for possible publication at the IEEE ICC 2013}

\begin{abstract}
%\boldmath
In this paper, a cooperative two-hop communication scheme, together with
opportunistic relaying (OR), is applied within a mobile wireless body area
network (WBAN). Its effectiveness in interference mitigation is investigated in a scenario where there are multiple closely-located networks. Due to a typical WBAN's nature, no coordination is used among different WBANs. A suitable time-division-multiple-access (TDMA) is adopted as both an intra-network and also an inter-network access scheme. Extensive on-body and off-body channel gain measurements are employed to gauge performance, which are overlaid to simulate a realistic WBAN working environment. It is found that opportunistic relaying is able to improve the signal-to-interference-and-noise ratio (SINR) threshold value at outage probability of 10\% by an average of 5 dB, and it is also shown that it can reduce level crossing rate (LCR) significantly at a low SINR threshold value. Furthermore, this scheme is more efficient when on-body channels fade less slowly.
\end{abstract}

\IEEEpeerreviewmaketitle

\section{Introduction}
Wireless Body Area Networks (WBANs) represent the next generation of personal area networks \cite{Astrin2007}. The elementary components of a typical WBAN are sensors and a gateway device, which is alternatively known as a hub. A traditional WBAN has a centralized topology, which is coordinated by the central hub. It has a round-robin direct data exchange behavior between each sensor and gateway node \cite{tg6_d}. This kind of system is widely used for patients monitoring and athletes' training \cite{Lewis2008}. There were approximately 11 million active units around the word in 2009, and this number is predicted to reach 420 million by 2014 \cite{ABIResearch}.

The pervasive use of WBAN thus increases the need for good coexistence between multiple WBANs. Imagine patients wearing such a system in a medical centre, the number of WBANs closely located is large in some periods of the day and they can move rapidly with respect to each other. Due to WBANs' nature of high mobility and potential large density, it is generally not feasible to have a global coordinator in such a circumstance.

Therefore, this leads to a challenging issue \-- interference between multiple closely located WBANs. Interference among WBANs is a major critical issue that can cause performance degradation and hence is a threat to reliable operation of any WBAN. A proposed method that will be investigated in this paper is to use two-hop cooperative communication with an opportunistic relaying scheme. Recently, two-hop cooperative communications, which is included as an option in the IEEE 802.15.6 standard \cite{tg6_d}, has been proved to overcome typical significant path loss experienced in single-link star topology WBAN communication \cite{CMDraft}. Several such cooperative communication schemes have been investigated for WBANs using either narrow-band \cite{Dong2011,SMITH:WCNC:2012,DErricoICC11,Ferrand:AT:2011} or ultra-wideband \cite{Chen09JSAC}. Its effectiveness in interference mitigation has also been studied using realistic on-body channel data and a simulated model of inter-WBAN channel data~\cite{Dong2012}. It has been shown in \cite{Dong2012} that the use of cooperative communication in any WBAN-of-interest is able to significantly mitigate interference by providing an up-to 12 dB improvement in SINR outage probability.

In this paper, the work is extended from \cite{Dong2012}, some similarities are:
\begin{enumerate}
\item{Intra- and inter-WBAN access scheme:}
time division multiple access(TDMA) across multiple WBANs is employed as well as within the intra-WBAN access scheme since it provides better interference mitigation with respect to power consumption and channel quality \cite{Zhang2010};
\item{Three-branch cooperative communication:}
A WBAN system in this paper uses two relay nodes to provide extra diversity gain at the receiver. Three branches are used with one direct link from sensor to gateway node and two additional links via two relays;
\item{Intra-WBAN channel model:}
Extensive on-body channel gain measurements are adopted as the inter-WBAN channel model for the analysis by simulation.
\end{enumerate}

There are also some major differences in our simulation setup:
\begin{enumerate}
\item{Diversity combining scheme:}
Instead of a three-branch selection combining (SC) scheme at the gateway node \cite{Dong2012}, an opportunistic relaying (OR) method is simulated. OR reduces complexity when compared with SC by adopting the concept that only a single relay with the best network path towards the destination forwards a packet per hop;
\item{Inter-WBAN channel model:}
A major contribution in this paper is that performance of cooperative WBANs using opportunistic relaying with no cooperation between WBANs is investigated with real inter-body channel gain measurements.
\end{enumerate}

In a multi-WBANs co-existence scenario, a WBAN's performance is more interference limited. Therefore, the effectiveness of the proposed scheme is analysed based on signal-to-interference-and-noise ratio (SINR). Since the duration of outage is more important than the probability of outage in analyzing the performances of a communication system with multiple co-channel interferers \cite{Annavajjala2010}, then here our analysis shows both the first order statistics of outage probability and second order statistics of level crossing rate (LCR) for SINR. Statistics are compared with respect to a traditional star topology WBAN with the same channel data employed.

%The rest of the paper is organized as follows. In Section II, the system model used is the simulation is explained. Section III simulates the WBANs co-existence using measured on-body and off-body channel data. Section III also provides analysis of results. Section IV provides some concluding remarks on coexistence between non-cooperative WBANs being better facilitated by cooperative communications on any given WBAN.

\section{System Model}

\subsection{Configuration of WBAN}
\begin{figure}
\centering
\includegraphics[width=0.6\columnwidth]{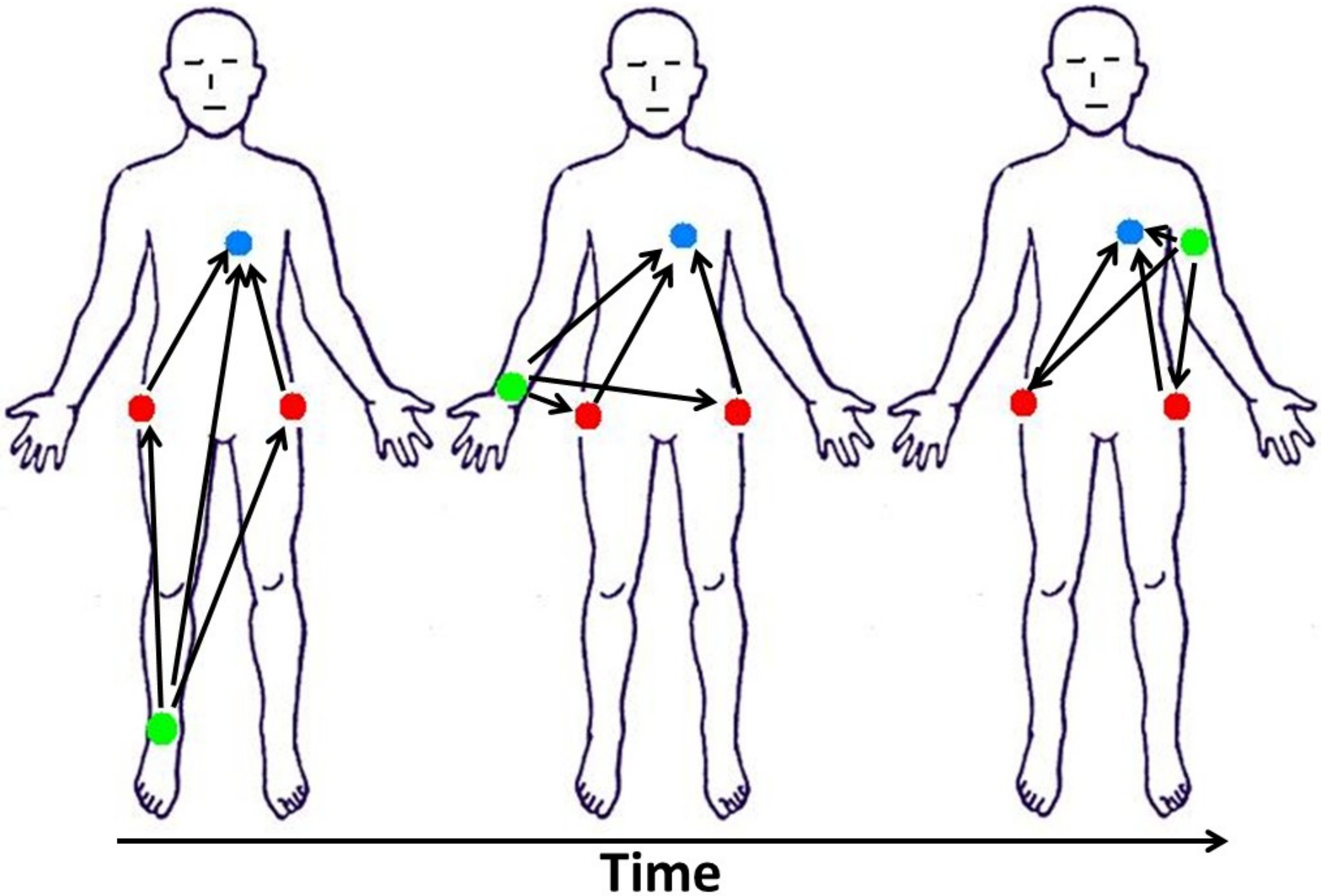}
\caption{WBAN configuration}
\label{fig: WBAN configuration}
\end{figure}

The model of a WBAN system used in the simulation consists of one hub (gateway node), two relay nodes and three sensor nodes, which are organized in a star topology. In order to cooperatively use measured channel data, the hub and two relays are placed separately at one of three locations: chest, left and right hips. For example, Fig. \ref{fig: WBAN configuration} shows the situation where hub is at chest and relays are at left and right hips respectively. In terms of sensor devices, they are located at any of the Rx places listed in Table \ref{node location}. Within each WBAN, sensor nodes are coordinated by the hub using a time division multiple access (TDMA) scheme. Therefore, as soon as the hub sends out a beacon signal to all connected sensors, they respond by transmitting required information back to hub in a pre-defined sequence during their allocated time slots. During the transmission period, one of the relays assists transmission by providing another copy of the signal to the hub. The choice of relay is based on an opportunistic relaying scheme that will be explained in a later section. In this paper, it is assumed that only one information packet is sent from each sensor during its allocated time slot. After completion of transmission from all sensor nodes, the system becomes idle until the next beacon period.

\subsection{Two WBANs coexistence}
For best co-channel interference mitigation and for reducing power consumption, TDMA is adopted as the co-channel access scheme across WBANs, as well as being the access scheme within any WBAN. Assume there are total $N_c$ WBANs co-located (or in close proximity) and the number is fixed during the period of simulation, then the shared channel can be evenly divided into $N_c$ time slots with a length of $T_d$. Therefore, each WBAN goes into idle status for a length of $(N_c-1)T_d$, which is also denoted as $T_{\mathrm{idle}}$, after it completes transmission. However, due to lack of a global coordinator among multiple WBANs, the execution of the TDMA scheme used here is slightly different from the traditional implementation of TDMA. Here, in our TDMA scheme the coordinator in each WBAN chooses the transmission time of every superframe randomly, following a uniform random distribution over $[0, T_d+T_{\mathrm{idle}}]$. In this paper, two WBANs employing this same configuration are used in analysis.

\subsection{intra- and inter-WBANs Channel Model}
\subsubsection{On-body channel data}
The extensive on-body WBAN channel data was measured with small wearable channel sounders operating at 2.36 GHz over several hours of normal everyday activity of an adult subject. This process was repeated several times with devices on different experiment subjects. The measuring system on a single subject consisted of three transceivers and 7 receivers. Their locations are listed in Table \ref{node location}. According to the experiment setup as shown in Table \ref{node location}, samples were taken over a period of two hours. During this process, three transceivers broadcasted in turn at 0~dBm in a round-robin fashion, with each one occupying the channel for 5 ms. Hence, each transceiver transmits every 15 ms. While one was transmitting, the remaining channel sounders recorded the received signal strength indictor (RSSI) if a packet was successfully detected.

\begin{table}[t!]
\centering
\caption{Tx/Rx Radio Locations, X indicates a channel measurement. LH-Left Hip, RH-Right Hip, C-Chest, HD-Head, RW-Right Wrist, LW-Left Wrist, LAR-Upper Left Arm, LA-Left Ankle, RA-Right Ankle, B-Back}
\begin{tabular}{|c|c|c|c|c|c|}
\hline
 & Rx-Hd & Rx-Rw & Rx-Lw & Rx-Lar & Rx-La\\ \hline
Tx-Lh & x & x & x & x & x\\ \hline
Tx-Rh & x & x & x & x & x\\ \hline
Tx-C & x & x & x & x & x\\ \hline
 & Rx-Ra & Rx-B & Rx-C & Rx-Lh & Rx-Rh\\ \hline
Tx-Lh & x & x & x & & x\\ \hline
Tx-Rh & x & x & x & x & \\ \hline
Tx-C & x & x & & x & x\\ \hline
\end{tabular}
\label{node location}
\end{table}

\subsubsection{Interfering channel data}
In order to study the impact of an opportunistic relaying scheme on a WBAN while multiple collocated WBANs work non-cooperatively, a well-defined and reliable inter-WBAN channel model is employed. NICTA has performed several measurements to capture such multi-person interference channel data. The experiment ran for many times with different number of subjects involved, each wearing channel sounders that were the same as for measuring on-body channel data. Each subject had a transceiver and two receivers, working at 2.36 GHz, attached to their body. In the experiment, the transceiver was at left hip and two receivers were at right upper arm and left wrist respectively. For the data set used in the simulation, there were total 8 people wearing such systems and their activities included walking together from office to a cafe and then back to the office. The transceiver on each subject transmitted packets at 0~dBm at 25~Hz in a round-robin fashion, i.e., each one occupies the channel sequentially for 5~ms. All units log the RSSI value from all packets received. Each device writes one output file when it interleaves the RSSI samples.

\subsubsection{Overlaying two data sets}
With these two data sets, a realistic multi-WBAN co-existence scenario can be simulated. However, we recall that the on-body channels and off-body interfering channels were sampled every 15 ms and 40 ms respectively. Therefore, in order to use these two data sets together, it is critical to synchronize them at the same sampling rate. Since the coherence time of both on-body and off-body channels is around 500~ms \cite{SMITH:WCNC:2012_2}, it is possible to down sample both types of channel gain coefficients to 120 ms per sample that is within the period while the channel is stable. In addition, a block fading model is used whereby the Tx-Rx channel gain is constant over each packet transmission period. Therefore, the time between each packet transmission is effectively 120~ms.

\begin{figure}[t!]
\centering
\includegraphics[width=0.6\columnwidth]{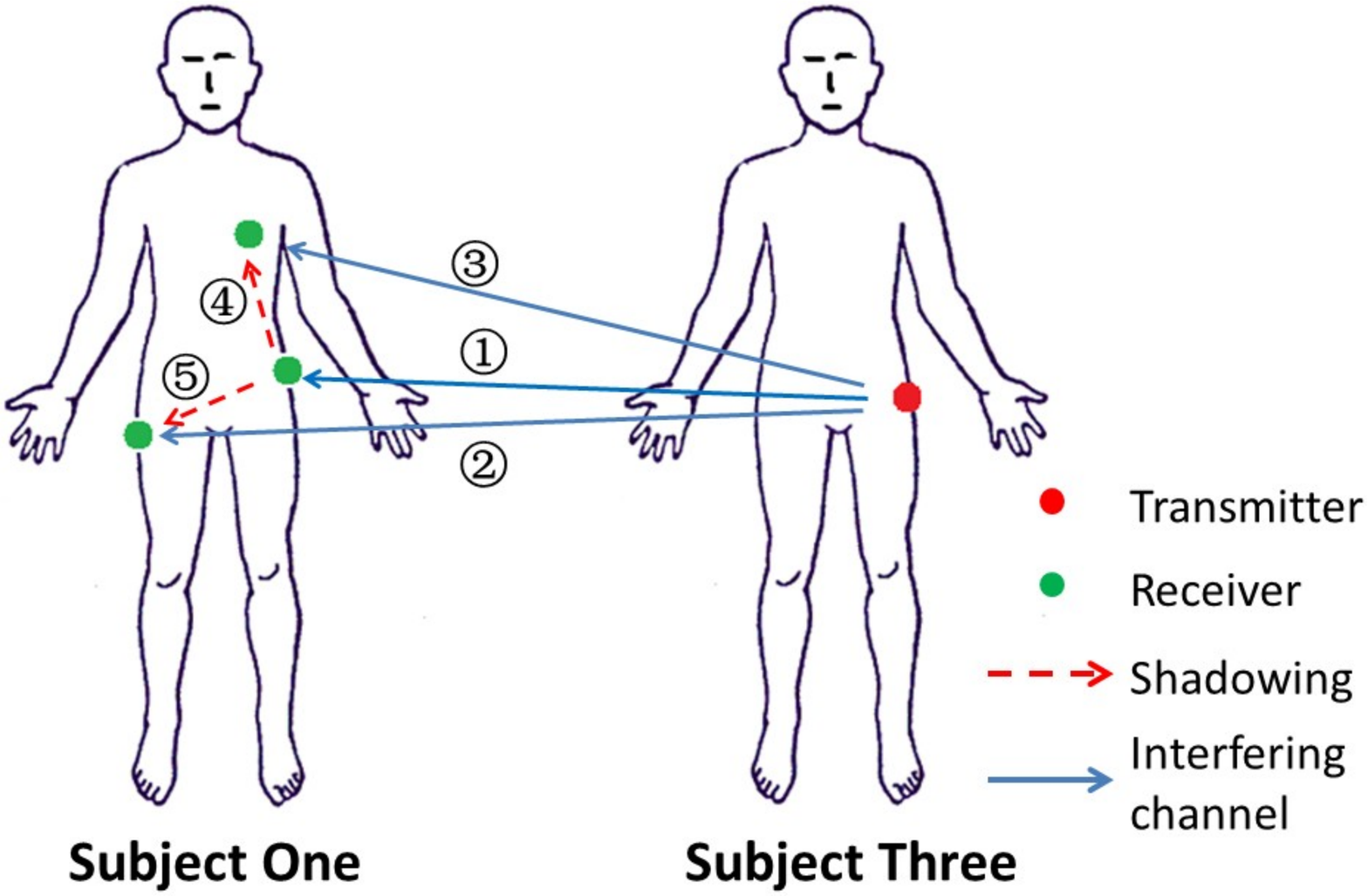}
\caption{Channel overlaying map}
\label{fig: Channel Overlaying map}
\end{figure}

Besides overlaying fading profiles in the time domain, it is also crucial to consider spatial overlaying. Assume the person-of-interest is named Subject One, and an interfering transmitter is attached to Subject Two. Gateway node and relays are placed on Subject One at the chest, left and right hips respectively (three common locations for these two types of devices). Since opportunistic relaying used in this paper is based on the SINR value at the relays, interferences are considered at three locations, i.e. one hub node and two relay nodes. Then these three devices on Subject One can all hear the interference from the interferer, i.e. Subject Two's transmitter, in which way the interfering system can influence the opportunistic relaying result of the WBAN-of-interest. Channel 1, 2 and 3 shown in Fig. \ref{fig: Channel Overlaying map} are the resultant interfering channels described previously. Based on this, the three interfering channels specified in Table \ref{Interfering channels} were used in simulation.

\begin{table}[t!]
\centering
\caption{Interfering channels}
\begin{tabular}{|c|c|c|}
\hline
Channel index & Interfering WBAN & WBAN-of-interest \\ \hline
1 & Left Hip & Left Hip \\ \hline
2 & Left hip & Right Hip \\ \hline
3 & Left Hip & Chest \\ \hline
\end{tabular}
\label{Interfering channels}
\end{table}

However, because the off-body channel data was measured with devices at left hip, right ankle and left wrist, there is no direct channel data for those mentioned in Table \ref{Interfering channels}. Thus, it is necessary to overlay two data sets spatially to form a good model of interfering channels between two subjects. Details of overlaying can be viewed from Table \ref{Channel composition/overlaying}. It can be seen that the shadowing components of Part 2 channels are combined with channel gain measurements listed in the Part 1 column. The shadowing components are abstracted by removing the corresponding free space path loss from the original channel gain data.

\begin{table}[t!]
\centering
\caption{In the table, (Intf) represents the person of interfering, and (I) indicates the person-of-interest. Numbers in each cell represent indicated channel in Fig.\ref{fig: Channel Overlaying map}}
\begin{tabular}{|l|l|l|}
\hline
Final channel & Part 1 & Part 2 \\ \hline
1: LH(Intf) \-- LH(I) & 1: LH(Intf) \-- LH(I) &  \\ \hline
2: LH(Intf) \-- RH(I) & 1: LH(Intf) \-- LH(I) & 5: LH(I) \-- RH(I) \\ \hline
3: LH(Intf) \--  C(I) & 1: LH(Intf) \-- LH(I) & 4: LH(I) \-- C(I) \\ \hline
\end{tabular}
\label{Channel composition/overlaying}
\end{table}

\subsection{Opportunistic Relaying Scheme}
% Description for Opportunistic relaying scheme needs to be paraphrased.!!
In the simulation, three-branch opportunistic relaying is adopted. Among three branches, one of them is the direct link from the active sensor to the gateway node, and the other two are decode-and-forward links via two relays, which are located at left and right hips respectively. The best relay is chosen prior to the sensor transmission and requires each relay to know the instantaneous signal strength it receives, as well as the instantaneous signal strength received from it at the hub. In this paper, it is achieved by choosing to maximize the minimum of the weighted channel strengths between sensor to relay and relay to hub links. Channel strength is quantified as SINR $\nu$, which is defined as follows:

%SINR equation
\begin{gather}
	\nu = \frac{a_{\mathrm{eff}}*|h_{TxRx}|^2}{|\varepsilon_{\mathrm{noise}}|^2 + \sum(a_{i}*|h_{\mathrm{int},i}|^2)}
\label{equ:SINR}
\end{gather}

where $a_{x}$ represents the transmit power of a signal packet and the subscripts $\mathrm{eff}$ and $\mathrm{int}$ indicate signal-of-interest and interference respectively; $\varepsilon_{\mathrm{noise}}$ is the instantaneous noise level at receiving node. $|h_{x}|$ represents average channel gain across the time duration of the transmitted signal packet, and its corresponding path is indicated as the subscripts $TxRx$ and $\mathrm{int},i$.

With opportunistic relaying, each decode-and-forward relay evaluates the minimal instantaneous received SINR among two hops in the beginning of a superframe. With SINR $\nu$ defined above, the selected relay $j_{OR}$ is hence determined by the following:

% Equation goes here
\begin{align}
j_{OR}& = \arg \max[ \nu_{1,min}, \nu_{2, min} ], \\
\nonumber \nu_{k,\min}& = \min[\nu_{r_k h}, \nu_{sr_k}],\:\textrm{with}\: k = 1\:\textrm{or}\:2.
\end{align}
The subscripts $s$, $r_k$ and $h$ represent sensor, $k$th relay and hub respectively; $\nu_{ab}$ represents the average SINR for the signal packet transmitted to $b$ from $a$.

At the receiving end, i.e. hub, the relayed copy is then compared with source-to-destination link. The decision is also based on the SINR value, and the larger one will be used for decoding and demodulation.

\section{Results}
Here, we simulate the case of a hub placed at chest with two relays at the left and right hips, following the simulation configuration described in Section II. We present performance analysis with respect to the system's first order statistics \-- outage probability of SINR, and its second order statistics \-- level crossing rate. Their concept will be defined in following sections. As shown in Table \ref{Simulation combination}, simulation is duplicated using different combinations of the person-of-interest and the interferer. Every combination repeats 10 times with a varied starting time for the channel data.

\begin{table}[t!]
\centering
\caption{Simulation combination: X indicates a simulation is performed.}
\begin{tabular}{|c|c|c|c|c|c|c|}
\hline
Subject Index & \#1 & \#2 & \#3 & \#4 & \#5 & \#6\\ \hline
\#1 & & x & x & x & x & x \\ \hline
\#2 & x & & x & x & x & x \\ \hline
\end{tabular}
\label{Simulation combination}
\end{table}

\subsection{Analysis of SINR outage probability}
Outage probabilities at given SINR thresholds are defined as the probability of SINR value being smaller than a given threshold
\begin{gather}
    Pr(\nu < \nu_{threshold})
\end{gather}
The performance of single link communication and two-hop cooperative communication schemes are compared with respect to SINR threshold at 1\% and 10\% outage probability. Here, 10\% outage probability corresponds to a guideline for 10\% maximum packet error rates in the IEEE 802.15.6 BAN standard \cite{tg6_d}.

\begin{figure}[]
\centering
\subfigure[Outage probability of Subject One with channel sample starting at n=45000]
{\includegraphics[width=0.96\columnwidth]{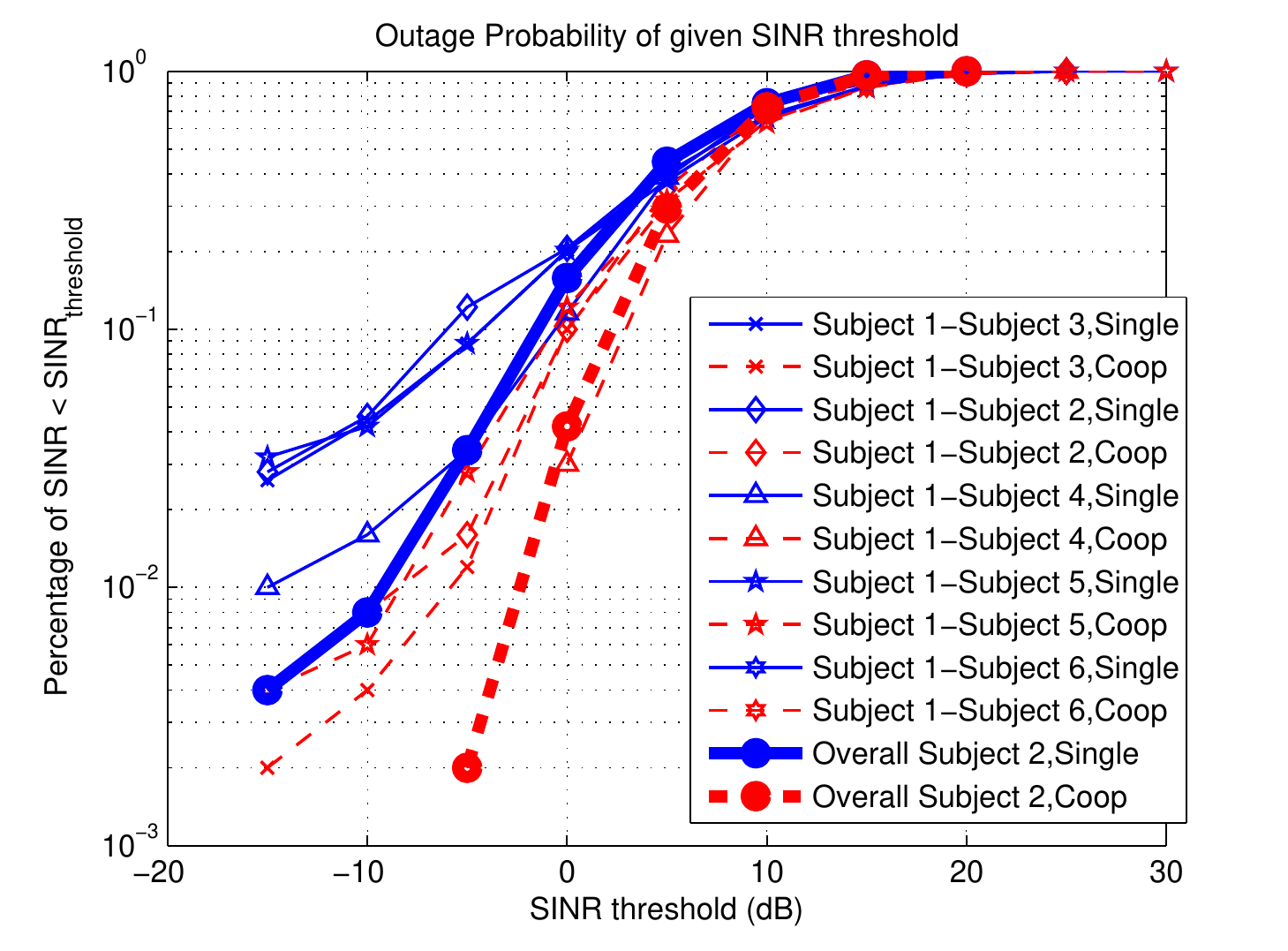}
\label{fig:subjectOne_outage}}
\subfigure[Outage probability of Subject Two with channel sample starting randomly]
{\includegraphics[width=0.96\columnwidth]{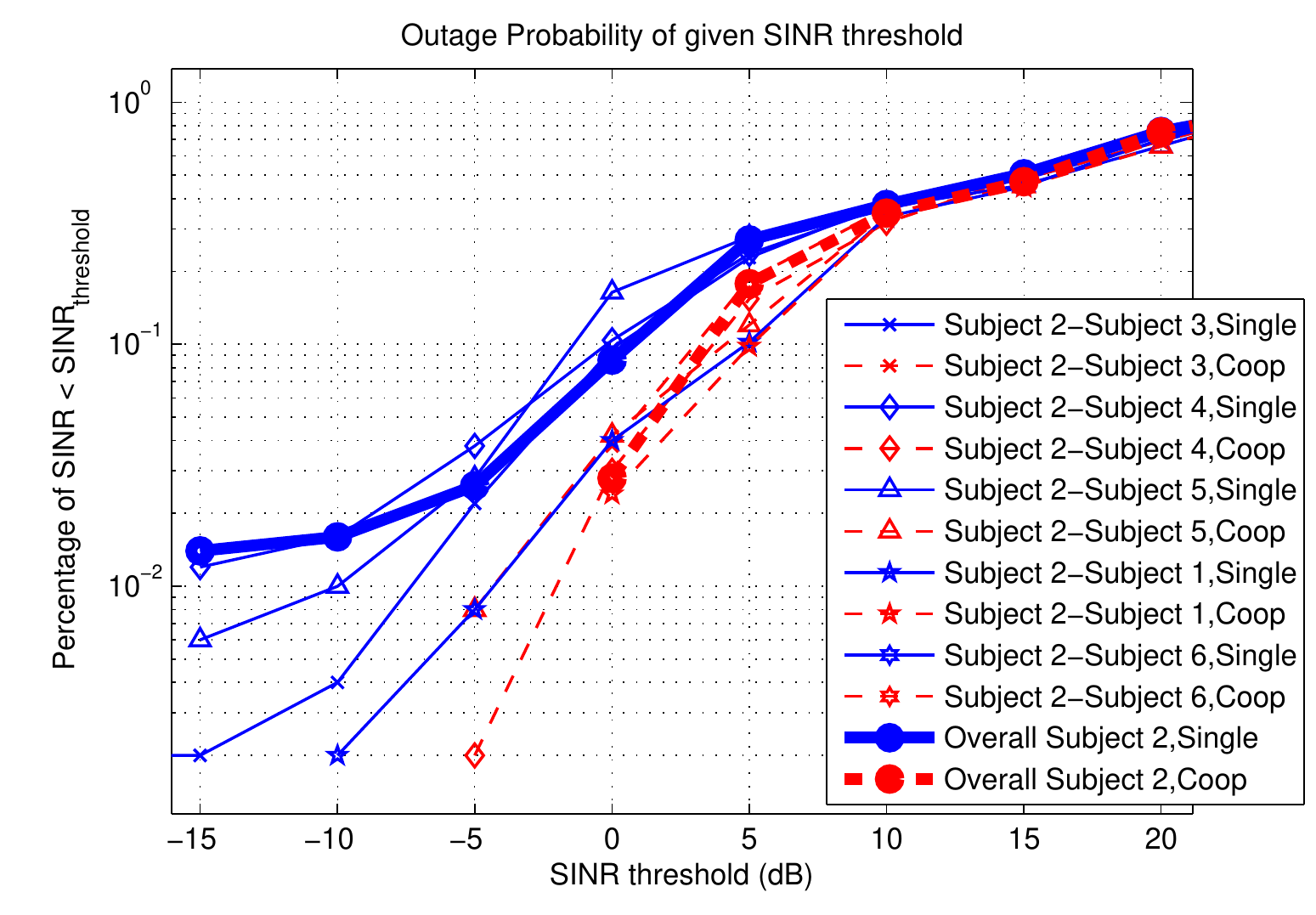}
\label{fig:subjectTwo_outage}}
\caption{Outage probability of SINR for two subjects}
\label{fig:outage probability}
\end{figure}

Fig. \ref{fig:outage probability} shows the outage probability for situations when different subjects are treated as the person-of-interest. Note that in Fig. \ref{fig:subjectOne_outage}, the relevance of choosing a starting sample index $n = 45000$ is explained in the next paragraph. For Subject One, the cooperative communication scheme provides about $5$~dB improvement over traditional single link communications at 10\% outage probability. At 1\% outage probability, the improvement is even more significant, there is, in fact, more than 15~dB improvement. In contrast, simulation on Subject Two shows similar results, with 4~dB and 6~dB improvement at 10\% and 1\% outage probabilities respectively.

\begin{figure}[]
\centering
\includegraphics[width=0.96\columnwidth]{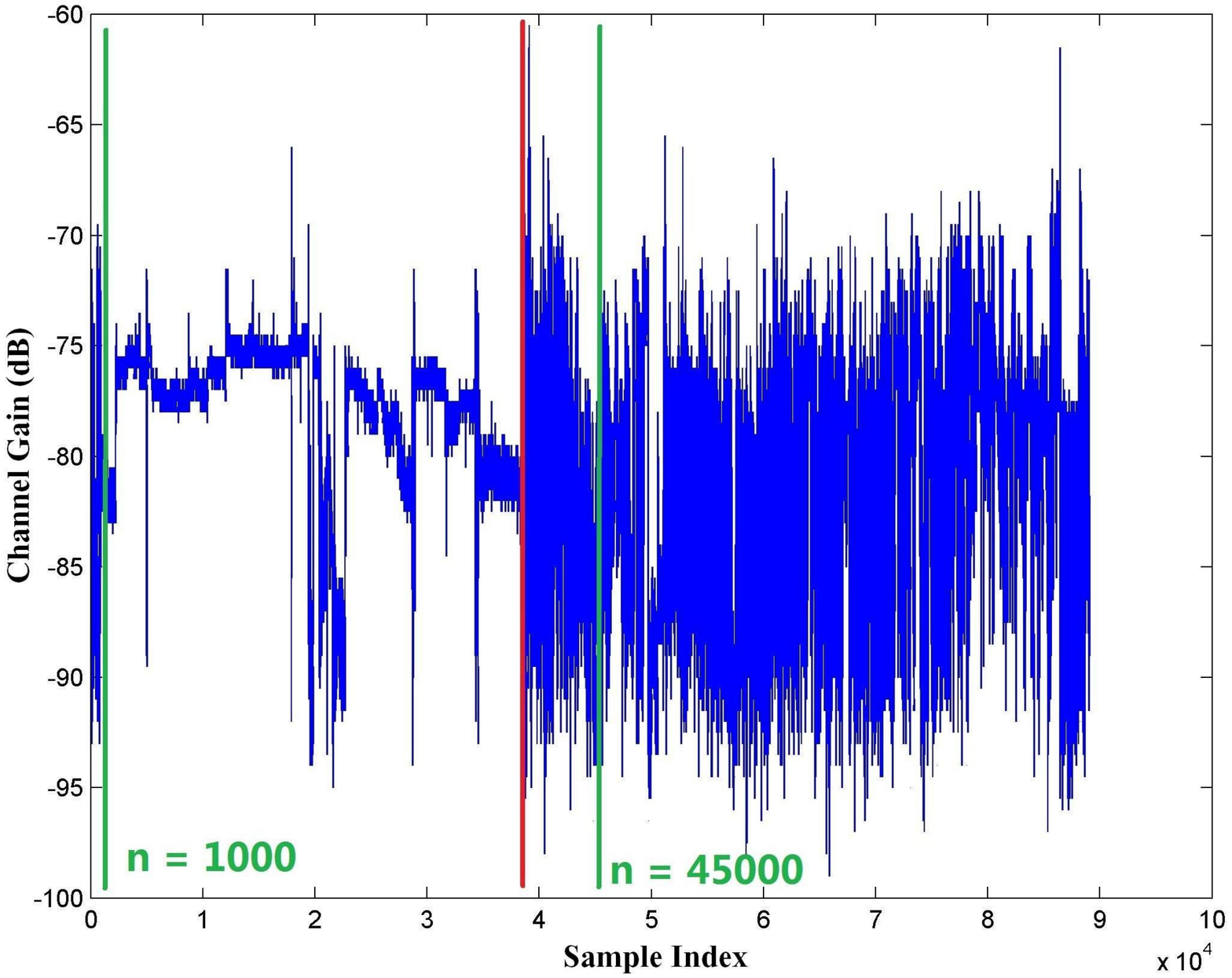}
\caption{A typical on-body channel gain plot of Subject One}
\label{fig: channel gain plot}
\end{figure}

\begin{figure}[]
\centering
\includegraphics[width=0.96\columnwidth]{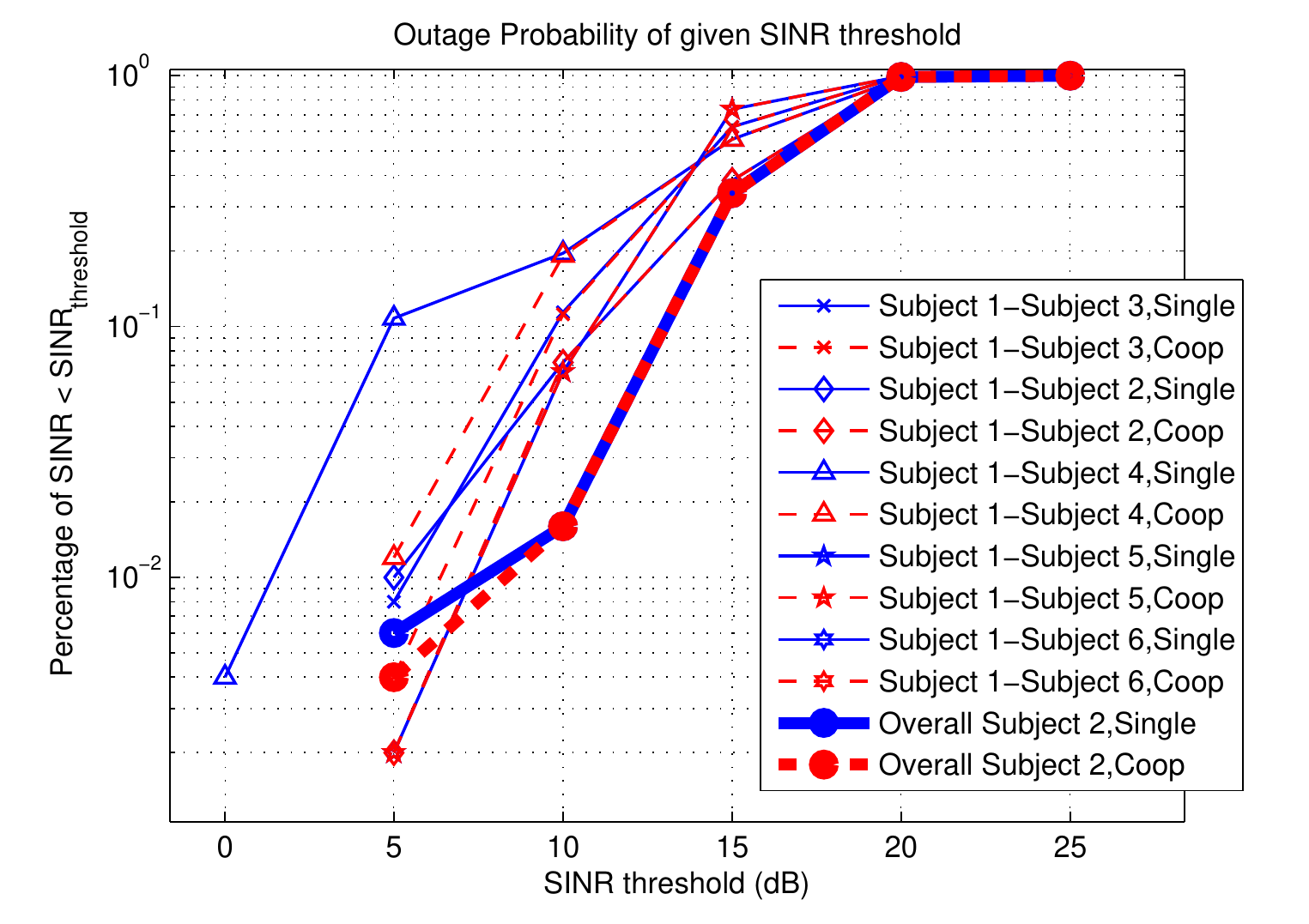}
\caption{Outage probability of Subject One with channel sample starting at t=0}
\label{fig: Outage probability of Subject One first half}
\end{figure}

However, while running simulations using a different starting channel sample index, it is found that channel stability has a significant impact on the performance of the cooperative communication scheme. Fig. \ref{fig: channel gain plot} is a typical on-body channel gain plot for Subject One. It is clear that there is a significant change in operating environment at the point where the red line is placed in Fig. \ref{fig: channel gain plot}. Subject One's on-body channels become more unstable after that point, i.e. channel coherence time decreases rapidly. Hence, simulation is also performed with channel sample taken from $n = 1000$, and result is plotted in Fig. \ref{fig: Outage probability of Subject One first half}. Compared to Fig. \ref{fig:subjectOne_outage} where channel sample is taken from $n = 45000$, cooperative communication scheme provides little SINR improvements at both 10\% and 1\% outage probabilities in this situation. As a result, two-hop cooperative communication schemes can provide even more improvement of the system performance when the environment in which the WBAN/s are operating changes rapidly.

\subsection{Level crossing rate}
Level Crossing Rate(LCR) of time-varying SINR is defined as the average frequency of a received packet's SINR value going below a given threshold in the positive direction. Assume there are a total of $n$ crossings at threshold $\nu_{\mathrm{threshold}}$, then the corresponding LCR value is calculated as
\begin{gather}
    LCR_{\nu_{\mathrm{threshold}}} = \frac{n}{\sum\limits_{i = 1}^{n-1}{t_{i,i+1}}}
\end{gather}
where $t_{i,i+1}$ is the time between the $i$th and $(i+1)$th crossing.

\begin{figure}[t!]
\centering
\includegraphics[width=0.96\columnwidth]{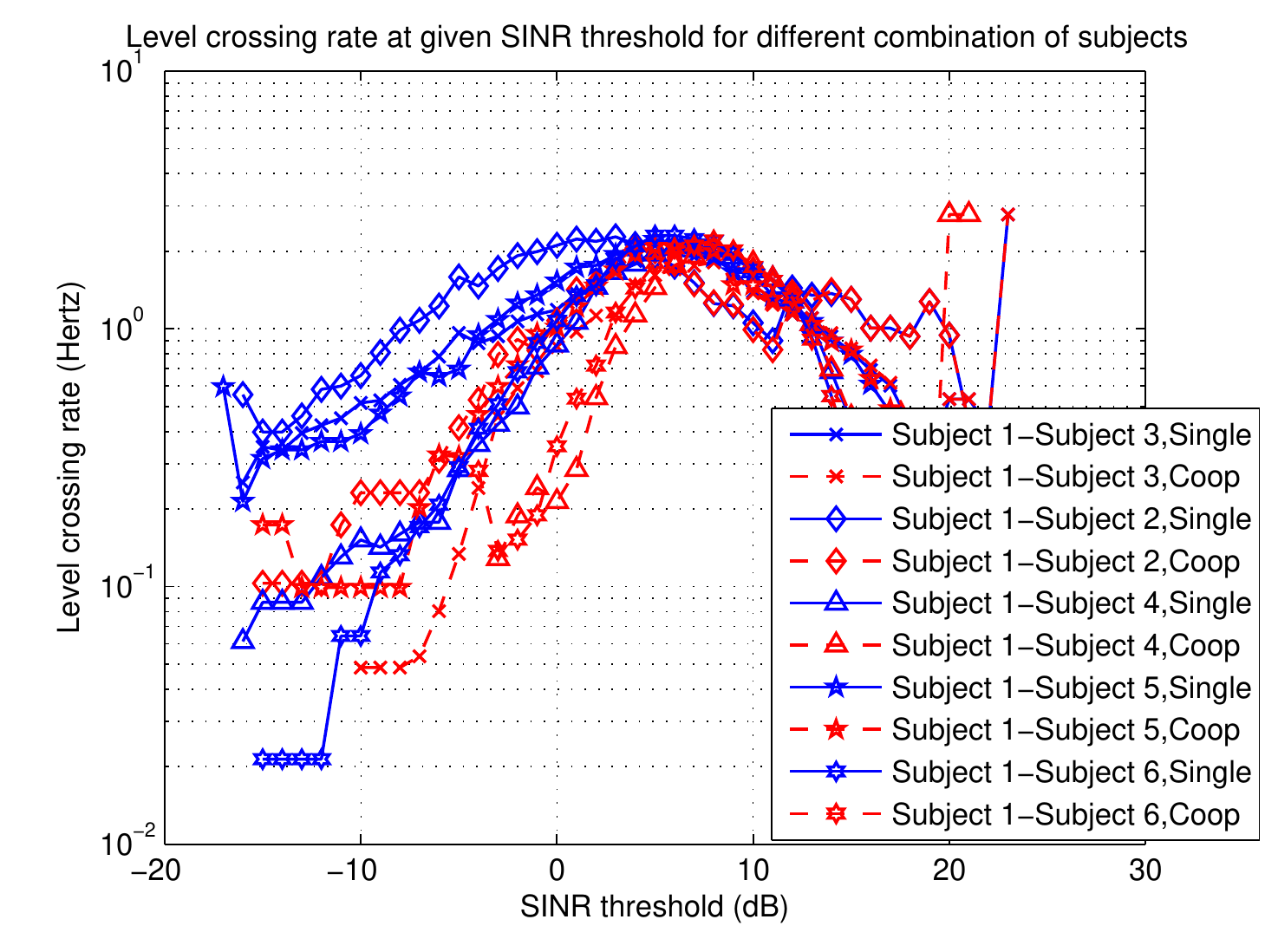}
\caption{LCR of Subject One with channel sample starting at t=45000}
\label{fig: LCR of Subject One second half}
\end{figure}

\begin{figure}[t!]
\centering
\includegraphics[width=0.96\columnwidth]{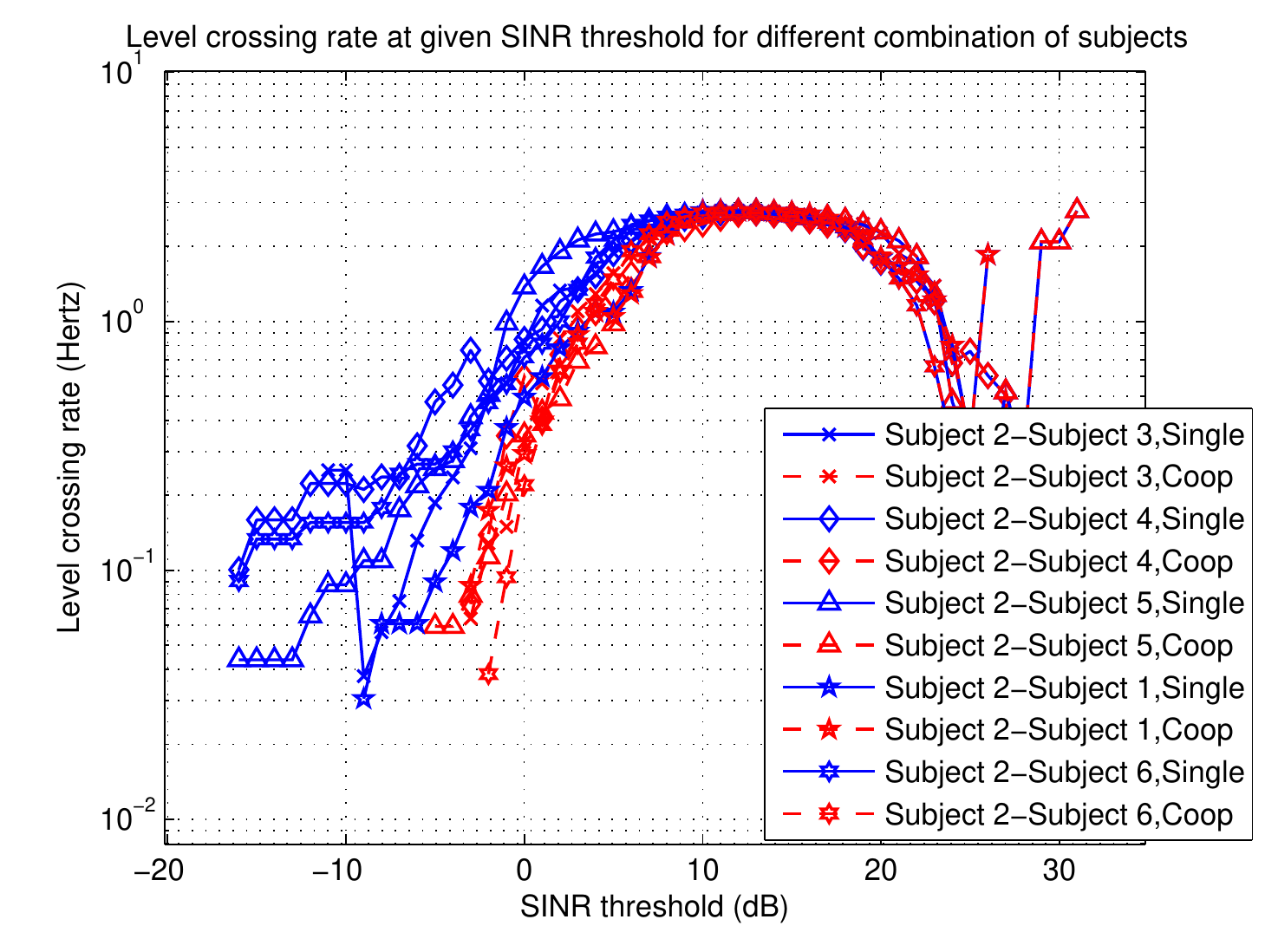}
\caption{LCR of Subject One with channel sample randomly}
\label{fig: LCR of Subject Two}
\end{figure}

\begin{figure}[t!]
\centering
\includegraphics[width=0.96\columnwidth]{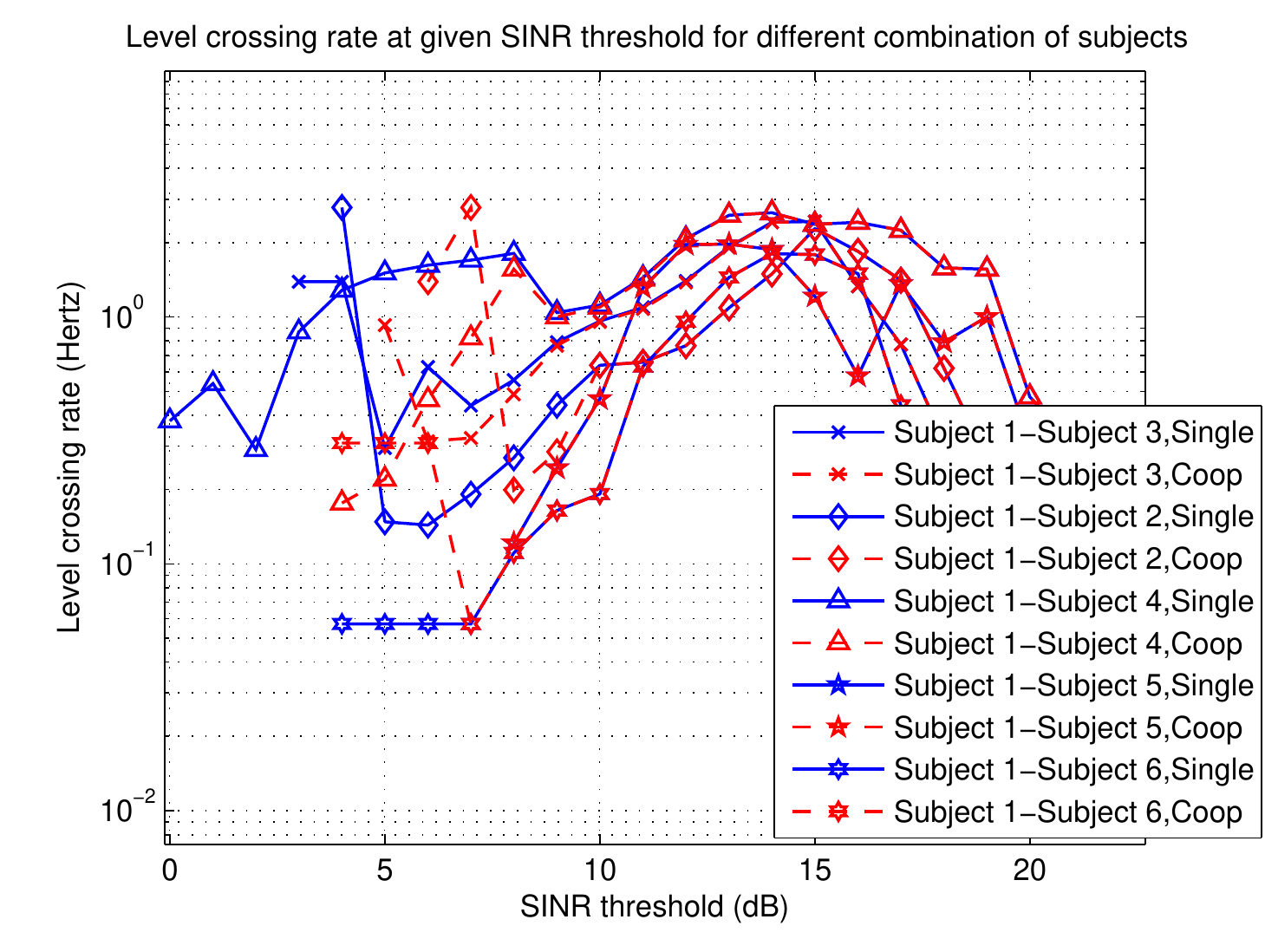}
\caption{LCR of Subject One with channel sample starting at t=0}
\label{fig: LCR of Subject One first half}
\end{figure}

Fig. \ref{fig: LCR of Subject One second half} and \ref{fig: LCR of Subject Two} show that cooperative two-hop communications are able to reduce level crossing rate at low SINR threshold values significantly. At LCR of 1~Hz, the SINR threshold value for Subject One rises by an average of 6~dB as shown in Fig. \ref{fig: LCR of Subject One second half} with the use of cooperative communications. For Subject Two, the improvement, as shown in Fig. \ref{fig: LCR of Subject Two}, is about 4~dB. System performance remains similar at high SINR threshold values. In addition, it can similarly be observed from Fig. \ref{fig: LCR of Subject One first half} in terms of outage probability, when coherence time of the on-body channels in the WBAN-of-interest is large (i.e., channels are more stable). In the case of Fig. \ref{fig: LCR of Subject One first half}, the LCR curves overlap most of the time for single-hop communications and cooperative communications schemes. Therefore, in such a case, there is no real performance advantage in terms of level crossing rate to use our proposed cooperative communication scheme.

\section{Conclusion}
In this paper, a three-branch opportunistic relaying scheme was investigated in a WBAN-of-interest under the circumstance where multiple WBANs co-exist non-cooperatively. TDMA was employed, as a suitable inter-network, as well as intra-network, multiple access scheme. Empirical inter-WBAN and intra-WBAN channel gain measurements were adopted to enable simulation of a practical working environment of a typical WBAN. Performance was evaluated based on outage probability and level crossing rate of received packets' SINR values.

It has been found that opportunistic relaying can provide an average of 5~dB improvement at an SINR outage probability of 10\%. It also reduces the level crossing rate significantly at low SINR threshold values. However, the performance of opportunistic relaying relies on the quality of the on-body channel greatly. Opportunistic relaying is particularly more useful, than single-link star topology communications, in a rapidly changing environment.

\bibliographystyle{IEEEtran}
%\bibliography{ICCref}

\end{document}